\documentclass[aps,prc,twocolumn,superscriptaddress,nofootinbib]{revtex4-1}
\usepackage{latexsym}
\usepackage{amsmath}
\usepackage{amssymb}
\usepackage{amsfonts}
\usepackage{bm}
\usepackage{physics}
\usepackage{booktabs}

\usepackage{color}
\definecolor{purple}{rgb}{0.5,0,0.5}
\definecolor{blue}{rgb}{0.0,0,0.9}
\definecolor{prdblue}{rgb}{0.133,0.118,0.498}
\usepackage[colorlinks=true, pdfstartview=FitV, linkcolor=prdblue, citecolor= prdblue, urlcolor=prdblue]{hyperref}

\usepackage{supertabular}
\usepackage{placeins}
\usepackage{epsfig}
\usepackage{graphicx}

\usepackage{CJKutf8}

\usepackage{soul}

\begin{document}
\begin{CJK*}{UTF8}{gbsn}

\title{A novel approach to proton-boron-11 fusion}

\author{Hong-Yi Wang (王弘毅)}
\affiliation{School of Physics, Nanjing University, Nanjing, Jiangsu 210093, China}

\author{Yu-Qi Li (李煜琦)}
\affiliation{School of Physics, Nanjing University, Nanjing, Jiangsu 210093, China}

\author{Qian Wu (吴迁)}
\affiliation{School of Physics, Nanjing University, Nanjing, Jiangsu 210093, China}

\author{Zhu-Fang Cui (崔著钫)}
\email[Contact author: ]{phycui@nju.edu.cn}
\affiliation{School of Physics, Nanjing University, Nanjing, Jiangsu 210093, China}

\date{\today}

\begin{abstract}
Proton-boron-11 (p-$^{11}$B) fusion is a highly attractive aneutronic pathway for clean energy production, offering abundant fuel, negligible neutron activation, and the potential for direct energy conversion of charged $\alpha$ particles. However, its practical implementation is severely hindered by the extremely high Coulomb barrier, necessitating ignition temperatures far beyond those of conventional deuterium-tritium reactions. In this work, we propose a novel approach to enhance the low-energy fusion cross-section by introducing a negative muon ($\mu$). Instead of relying on the thermal equilibrium formation of a muonic molecule, we investigate a kinetic scenario in which a muonic hydrogen atom (p$\mu$) is formed first and subsequently bombarded with a $^{11}$B nucleus. We quantitatively characterize the dynamic screening of the proton's Coulomb field by the tightly bound $\mu$ cloud, the resulting modified Coulomb potential substantially lowers the effective barrier at intermediate separations. We also evaluate the penetrability, reaction cross-section, and reactivity of the p$\mu$-$^{11}$B system, the results indicate that the inclusion of $\mu$ enhances the tunneling probability by several orders of magnitude at incident energies below 100~keV, thereby significantly reducing the threshold for the nuclear reaction. This mechanism offers a promising alternative perspective for catalyzing p-$^{11}$B fusion, and also suggests a potential ignition pathway.
\end{abstract}

\maketitle
\end{CJK*}



\section{Introduction}
Life on Earth would not be possible without the nuclear fusion processes that power the Sun and other stars, where light-weight atoms are compressed under extreme pressure and heat to form heavier atoms with the help of gravity, produce clean and nearly limitless energy. Unlike fission power plants that produce radioactive waste which can remain dangerous for thousands of years, fusion products are usually harmless or only short-lived waste, while the energy efficiency is expected to be $\sim$4 times better than the former. Another important feature is that fusion reactions would automatically stop if the plasma falls below a certain temperature or density, so that safety is not a problem. For these reasons, researchers have been trying to make fusion reactors, in other words, `artificial suns' on Earth, work since the 1950s -- a 70-year dream and true triumph for science, engineering and society. Although so far reactions haven’t been able to keep running long enough to produce more energy than it takes to spark them, it is generally thought that practical nuclear fusion finally seems to be on the horizon~\cite{Barbarino2020,x2021,Jenko2025,FFDB2025,IAEA2025}.

Of course, there remains many things to be done. For example, one reason that makes it so difficult to conduct in a controlled manner is the challenge of containing electrically charged plasma at temperatures much hotter than the center of the Sun. Related, low energy nuclear reactions (LENR), also known as chemically assisted nuclear reactions (CANR) and cold fusion, which can also impact other areas of science and technology, has draw special attentions and is worthy of engagement from the broader scientific community~\cite{Rafelski1987,BERTULANI2016,Berlinguette2019,Chen2025}. Compared with other attempts, this might present solutions that require shorter timelines or less extensive infrastructure.

Regarding fuels, nowadays many reactors are fusing deuterium (D) with tritium (T), which is the easiest way; unfortunately, T is rare and this reaction produces neutrons, which can make the chamber radioactive. In contrast, proton with boron-11 (p-$^{11}$B) is an attractive aneutronic fusion pathway that has garnered significant attention in clean energy research in recent years~\cite{Margarone2022,Liu2023,Magee2023,Lindsey2024,Liu2024,Ochs2024,Ogawa_2024,Nissim2025,Sciscio2025,Liu2025,Hwang_2025,Liu_2026,wang2026}. The dominant reaction, p + $^{11}$B $\rightarrow$ 3$\alpha$ + 8.7 MeV, completely converts reactants into three charged $\alpha$ ($^{4}$He) particles. This characteristic offers distinct advantages: firstly, the fuels p (or hydrogen H) and $^{11}$B are abundant and stable in nature, presenting no radioactive management challenges; secondly, the neutron-free reaction products significantly reduce material activation, avoiding the complex nuclear waste handling issues inherent in traditional D-T and other fusions; furthermore, the energy of the charged $\alpha$ particles produced can be directly converted into electricity, offering a novel pathway for efficient fusion energy utilization. However, practical application of p-$^{11}$B fusion faces severe challenges. The reaction cross-section is relatively small in the low-energy region, requiring extremely high ion temperatures (approximately 300 keV) to achieve ignition. Therefore, exploring practically viable methods to effectively enhance the reaction cross-section of p-$^{11}$B has become a crucial topic. 

As a promising way of LENR/CANR, theoretical and experimental studies have demonstrated that the introduction of muons can significantly enhance the fusion reaction cross-sections, known as muon-catalyzed fusion (MCF or $\mu$CF)~\cite{Frank1947,Jones1986,Breunlich1989,Ponomarev1990,Froelich1992,PhysRevLett.85.1642,PhysRevC.107.034607,Wu2024,Wu:2025bnm}. Since the muon mass is approximately 207 times that of the electron and also carries one unit of charge, the Bohr radius of the muonic atom is about 207 times smaller than that of a conventional electronic atom. This causes the muon to orbit the nucleus much more tightly, providing better screening of the nuclear charge, which originally carries one unit of positive charge, thereby substantially reducing the hindrance of the Coulomb repulsive potential.

Unfortunately, although the physics of D-T MCF is relatively well established, the introduction of boron ($\mathbb{Z}$=5) fundamentally alters the landscape, since the screening efficiency of the muon and the probability of capture are heavily dependent on the atomic number $\mathbb{Z}$~\cite{Koshigiri1984,PhysRevC.65.025503}. The high nuclear charge of boron creates a ``muon trap'', where the muon would tightly bound to the boron nucleus, significantly reducing its orbital radius and its ability to screen a second nucleus, the proton~\cite{Schaller1993}. This creates extreme physical barriers to both molecular formation and catalytic cycling, rendering standard thermal MCF mechanisms largely ineffective. In this study, we investigate the possibility that a muon and a proton forms a muonic hydrogen atom p$\mu$ first, then fuse with boron. We first calculate the effective charge experienced by the incident $^{11}$B as a function of the relative distance, quantitatively characterizing the screening effect of the muon on the proton's charge. Subsequently, the new Coulomb potential energy for the p$\mu$-$^{11}$B system and the function describing the minimum distance $r_{\text{min}}$ achievable between p and $^{11}$B as a function of the incident energy of $^{11}$B are evaluated. 

This paper is organized as follows. Sec.~\ref{pbmcf} presents a possible role of muon to accelerate the p-$^{11}$B reaction. Sec.~\ref{results} gives the results and discussion. Conclusions are given in Sec.~\ref{conclusion}.


\section{Muons and proton-boron-11 fusion}
\label{pbmcf}
In this section, we examine the potential scenario in which a muon and a proton initially form a muonic hydrogen atom p$\mu$, which subsequently fuses with boron. With the developments of muon production techniques worldwide, this may work as a possible way of LENR/CANR; on the other hand, this can also serve as a key to spark the p-$^{11}$B and similar systems, since, as argued in Refs.~\cite{Labaune:2013dla, Sciscio2025}, at high temperatures, the equilibrium thermal fusion rate of p-$^{11}$B is comparable to the D-T fusion rate.

\subsection{Muon Charge Screening}
\begin{figure}
	\centering
	\includegraphics[width=\linewidth]{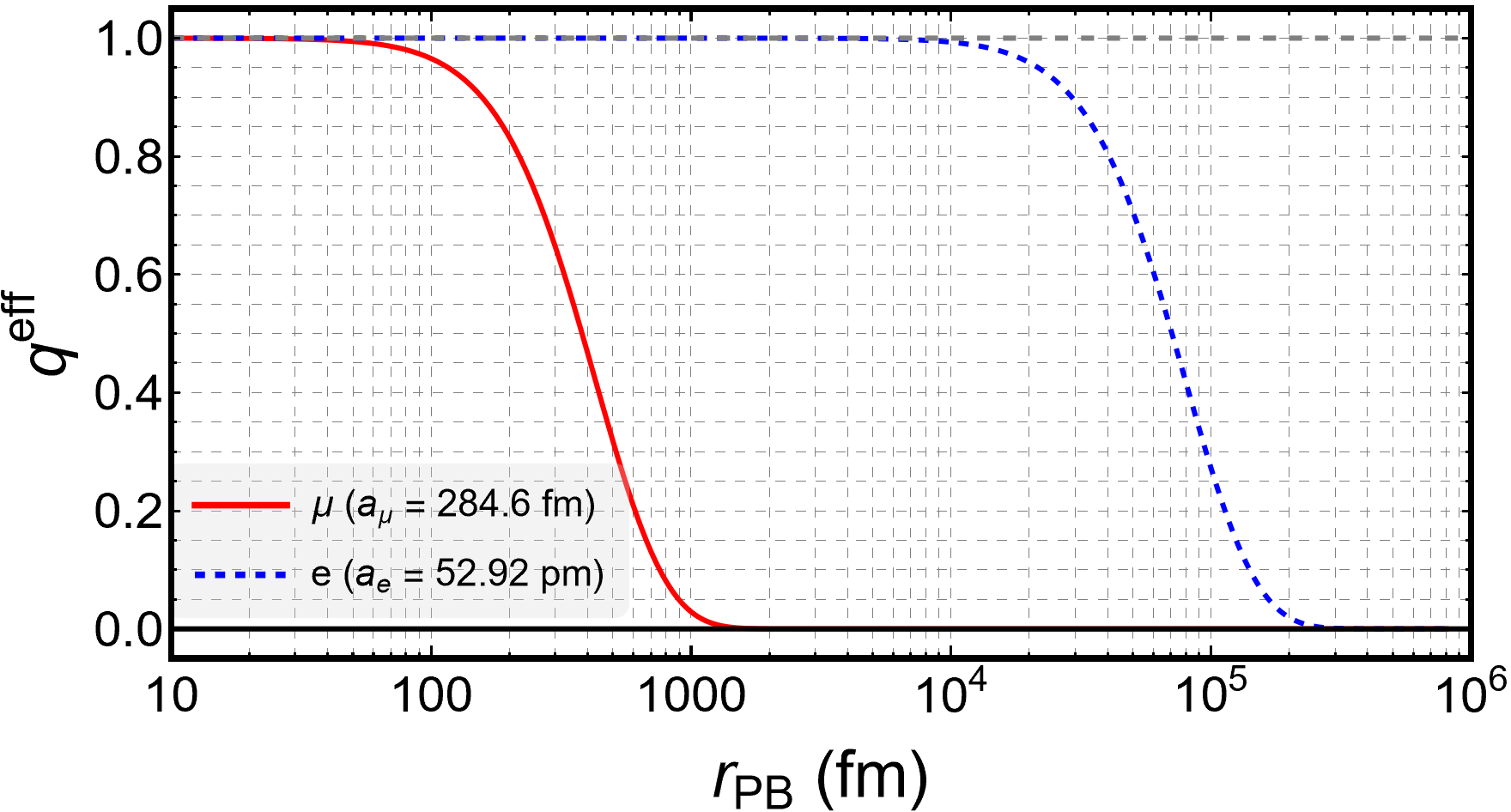}
    \caption{(Color online) 
The total effective charge experienced by $^{11}$B at a separation $r_{\rm pB}$ from the proton. The red solid and blue dashed lines correspond to muonic and electronic hydrogen respectively.
} 
    \label{qeff}
\end{figure}

Since we consider charged $^{11}$B accelerated to bombard the electrically neutral p$\mu$ target, we assume the proton is stationary, and the $^{11}$B nucleus possesses an initial kinetic energy relative to the proton, while the muon forms a muon cloud around the proton, screening its positive charge. We suppose the muon as being in a bound stationary state (static $1s$ wavefunction), neglecting dynamic polarization or transitions of the muonic wavefunction induced by the incident $^{11}$B nucleus. Therefore, the muonic wavefunction in a hydrogen-like $1s$ state is written as,
\begin{equation}
\psi(r)=\frac{1}{\sqrt{\pi a_{\mu}^{3}}} e^{-r / a_{\mu}},
\end{equation}
where $a_\mu$ is the first Bohr radius of the muon,
\begin{equation}
a_{\mu}=\frac{4 \pi \varepsilon_{0} \hbar^{2}}{m_{\mu} e^{2}} \approx 284.6\ \text{fm}.
\end{equation}
For a spherically symmetric, uniformly distributed muon cloud, the electric field felt by the $^{11}$B nucleus is equivalent to that from a point charge located at the proton, with a magnitude equal to the charge of the muon cloud contained within a sphere of radius $r_{\mathrm{pB}}$\cite{Jackson1998},
\begin{equation}
\begin{aligned}
q_{\mu}^{\mathrm{eff}} & = q_{\mu} \int_{0}^{r_{\mathrm{pB}}}|\psi(\mathbf{r})|^{2} d^{3} r \\
& = q_{\mu}\left[1-\left(1+\frac{2 r_{\mathrm{pB}}}{a_{\mu}}+\frac{2 r_{\mathrm{pB}}^{2}}{a_{\mu}^{2}}\right) \exp \left(-\frac{2 r_{\mathrm{pB}}}{a_{\mu}}\right)\right].
\end{aligned}
\end{equation}
Thereby the total effective charge felt by $^{11}$B at a distance $r_{\mathrm{pB}}$ from the proton, as shown in Fig.~\ref{qeff}, is,
\begin{equation}
q_{\mathrm{eff}}=q_{\mathrm{P}}+q_{\mu}^{\mathrm{eff}} = -q_{\mu}\left(1+\frac{2 r_{\mathrm{pB}}}{a_{\mu}}+\frac{2 r_{\mathrm{pB}}^{2}}{a_{\mu}^{2}}\right) \exp \left(-\frac{2 r_{\mathrm{pB}}}{a_{\mu}}\right).
\end{equation}

From Fig.~\ref{qeff}, we observe the variation of the effective charge of the $\mu$p atom and the ordinary hydrogen atom with distance. The screening of the central positive charge by the muon is significantly enhanced. At the same effective charge value, the muon reduces the distance by over two orders of magnitude, approximately equal to the mass ratio of the muon to the electron $\approx$ 207.
Thus, under the influence of this effective charge, the Coulomb potential energy for the p$\mu$-$^{11}$B system can be written as:
\begin{equation}
V_{\mathrm{P\mu-B}}^{\mathrm{eff}}\left(r_{\mathrm{pB}}\right)=\frac{1}{4 \pi \varepsilon_{0}} \frac{q_{B} q_{\mathrm{eff}}\left(r_{\mathrm{pB}}\right)}{r_{\mathrm{pB}}}.
\end{equation}
As the $^{11}$B nucleus radially approaches the proton, the net charge of the system transitions from ``neutral'' at large distances to that of bare nuclei at close range. The muon provides partial shielding of the Coulomb potential at intermediate scales, thereby lowering the effective barrier. We expect this effect to enhance the penetration probability when nuclear reactions occur at close distances.

\subsection{Penetrability}

For the calculation of the tunneling probability, we employ the WKB approximation. First, we use a magnitude criterion based on the action to examine the applicability range of the WKB approximation. The action is calculated as follows:
\begin{equation}
S \equiv \frac{1}{\hbar} \int_{r_{n}}^{r_{\min }(E)} \sqrt{2 m\left[V_{\rm pB}(r)-V_{\rm pB}\left(r_{\min }\right)\right]} d r,
\end{equation}
here $r_{\text{min}}$ is identified as the classical turning point, and assuming that from this point $^{11}$B may undergo instantaneous tunneling into the nuclear force range to fuse with the proton. 
\begin{figure}
	\centering
	\includegraphics[width=\linewidth]{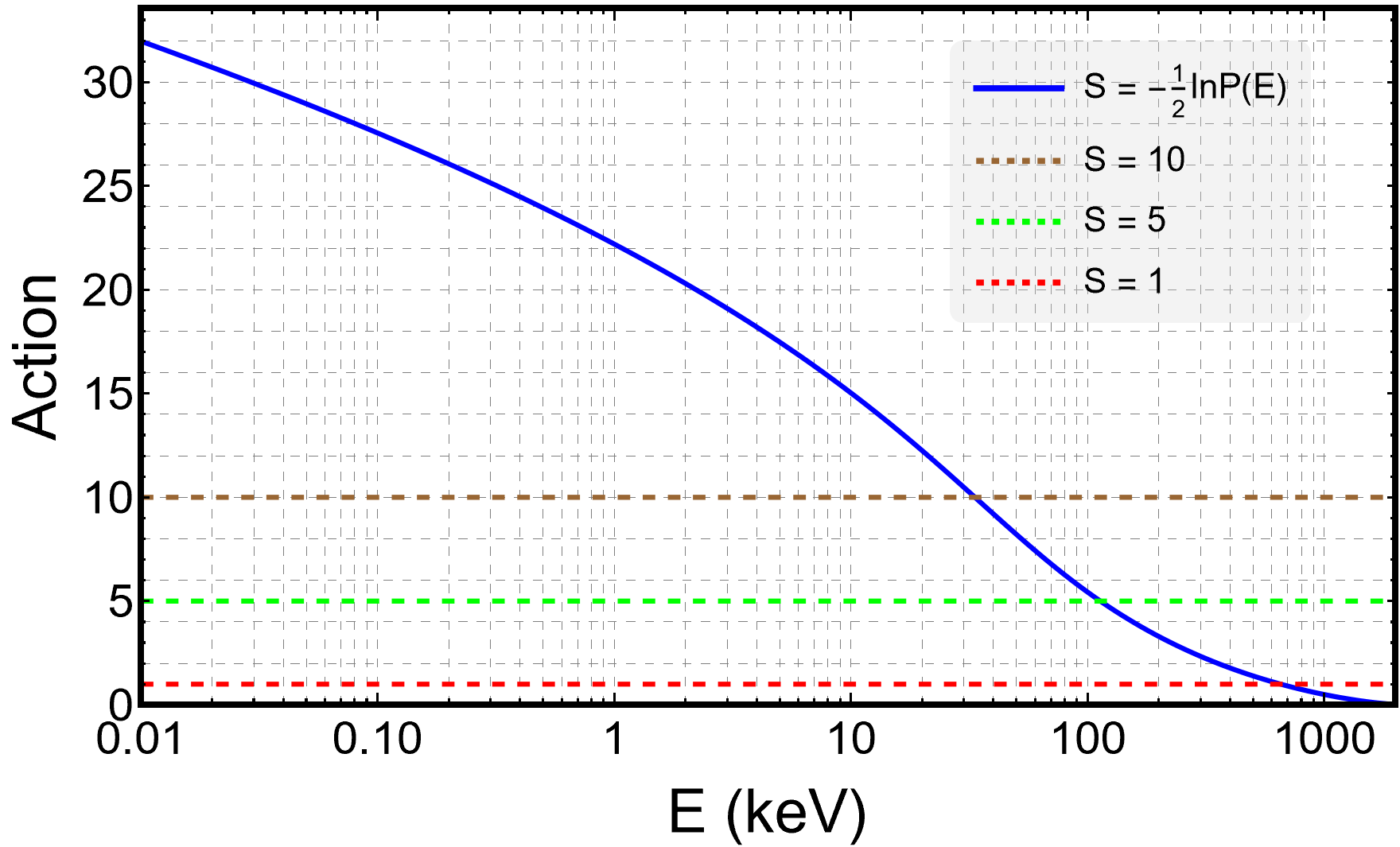}
    \caption{(Color online) Action $S$ as a function of incident energy $E$. The blue solid curve shows the action calculated using Eq. (6), which decreases monotonically with increasing energy. The horizontal brown, green, and red dashed lines indicate $S = 10$, $5$, and $1$, respectively, marking the reliability boundaries of the WKB approximation. According to the criterion adopted in this work, WKB is considered reliable for $S \geq 10$ (i.e., $E \leq 33.50$ keV), completely invalid for $S < 1$ ($E \geq 658.86$ keV), and of moderate reliability in the intermediate region. The Airy approximation is employed for energies above $33.50$ keV to connect with the WKB result at the turning point.} \label{fig:SvsE}
\end{figure}

The WKB approximation requires the action $S \gg 1$. However, in practical applications, this condition is difficult to satisfy. Following many previous works, we consider the WKB approximation to yield reliable results when $S \geq 10$, and we deem it to fail when $S < 1$. We compute the $S$ value for the $\mu$p-$^{11}$B reaction, as shown in Fig.~\ref{fig:SvsE}. It can be seen that $S$ decreases monotonically with increasing incident energy of $^{11}$B. The value of $S$ drops to 10 at energy $E = 33.50$ keV, to 5 at $E = 113.16$ keV, and to 1 at $E = 658.86$ keV. Therefore, for $E < 33.5$ keV, the WKB approximation can provide convincing results; for $E \geq 658.86$ keV, it fails completely; in the interval $33.5 - 113.16$ keV, its reliability is moderate. We will use the WKB approximation in the energy region below $33.5$ keV, and employ the Airy approximation starting from $33.5$ keV, which is a standard method to connect to the WKB result at the turning point~\cite{LandauLifshitz1977,Olver1974}.
We write the radial Schrödinger equation for p-$^{11}$B:
\begin{equation}
\frac{d^{2} u}{d r^{2}}+k^{2}(r) u(r)=0, \quad k^{2}(r) \equiv \frac{2 \mu}{\hbar^{2}}[E-V(r)].
\end{equation}
The approximate WKB wavefunction solution is then~\cite{LandauLifshitz1977}:
\begin{equation}
u_{\text {WKB}}(r) = \frac{C}{\sqrt{|k(r)|}} \exp \left(-\int_{r_{\mathrm{n}}}^{r}\left|k\left(r^{\prime}\right)\right| d r^{\prime}\right),
\end{equation}
where $C$ is a normalization constant. The WKB tunneling probability is thus obtained as:
\begin{equation}
P_{\mathrm{WKB}}(E)=\exp \left\{-\frac{2}{\hbar} \int_{r_{\mathrm{n}}}^{r_{\min}} \sqrt{2 m\left[V_{\mathrm{pB}}(r)-V_{\mathrm{pB}}\left(r_{\min}(E)\right)\right]} d r\right\},
\end{equation}
where $V_{\mathrm{pB}}(r)$ is the Coulomb potential energy between the proton and $^{11}$B, and $r_{\mathrm{n}}=3.284$ fm \cite{IAEANuclearData} is the sum of the nuclear radii of p and $^{11}$B.
In the region where the WKB approximation fails, the Airy approximation is typically a good method for connection~\cite{AbramowitzStegun1964,Olver1974}. At the chosen turning point $r_t = r_{\min}(E=33.5\ \text{keV})$ where the connection is made, the Coulomb potential is linearized: $V(r) \simeq E + V'(r_t)(r - r_t)$. The Airy variable is introduced as:
\begin{equation}
\xi=\kappa\left(r-r_{t}\right), \quad \kappa=\left(\frac{2 \mu V^{\prime}\left(r_{t}\right)}{\hbar^{2}}\right)^{1 / 3},
\end{equation}
transforming the Schrödinger equation into the standard Airy equation, whose general solution is: $u_{\text{Airy}}(\xi)=A \text{Ai}(\xi)+B \text{Bi}(\xi)$, where $\text{Ai}$ and $\text{Bi}$ are the two kinds of Airy functions. The coefficients $A$ and $B$ are determined by matching the Airy solution and its derivative to the WKB wavefunction and its derivative at a matching point $r_m$:
\begin{equation}
\left[\begin{array}{cc}\operatorname{Ai}\left(\xi_{m}\right) & \operatorname{Bi}\left(\xi_{m}\right) \\ \kappa \operatorname{Ai}^{\prime}\left(\xi_{m}\right) & \kappa \operatorname{Bi}^{\prime}\left(\xi_{m}\right)\end{array}\right]\left[\begin{array}{l}A \\ B\end{array}\right]=\left[\begin{array}{c}u_{\mathrm{WKB}}\left(r_{m}\right) \\ u_{\mathrm{WKB}}^{\prime}\left(r_{m}\right)\end{array}\right].
\end{equation}
We choose $r_m$ as where $\left|\xi_{m}\right|=1$. At this point, both WKB and Airy approximations are valid, ensuring they describe the same physical solution.

\subsection{Reaction Cross-Section and Reaction Rate}
The reaction cross-section is an important bridge connecting experiment and theoretical calculation. Experiments cannot directly measure the probability of tunneling but instead measure reaction cross-section data. The scattering cross-section can typically be written in the following Gamow form:
\begin{equation}
\sigma(E)=\frac{S(E)}{E} P(E),
\end{equation}
where $1/E$ characterizes the influence of the geometric effect of the wavefunction on the reaction probability. The astrophysical S-factor $S(E)$ characterizes the effect of nuclear interactions on the reaction probability. $P(E)$ characterizes the tunneling probability for the nuclei to overcome the Coulomb barrier.
In thermonuclear fusion systems, ignition constraints and fusion power density at a fixed plasma pressure are calculated based on the reaction rate. Therefore, whether assessing the ignition conditions for p-$^{11}$B or arguing for its feasibility as a future fusion fuel, accurately understanding the p-$^{11}$B reaction rate at different temperatures is crucial. The calculation of the reaction rate is highly dependent on the scattering cross-section, with the expression given by:
\begin{equation}
\langle\sigma v\rangle=\frac{(8 / \pi)^{1 / 2}}{\mu^{1 / 2}\left(k_{B} T\right)^{3 / 2}} \int_{0}^{\infty} \sigma E \exp \left(-\mathrm{E} / \mathrm{k}_{\mathrm{B}} \mathrm{T}\right) \mathrm{dE},
\end{equation}
where $\mu$ is the reduced mass of p and $^{11}$B, $k_B$ is the Boltzmann constant, $T$ is the temperature, and $\sigma$ is the total scattering cross-section of the reaction.


\section{Results and discussion}
\label{results}
\subsection{Minimum Distance}

\begin{figure}
	\centering
	\includegraphics[width=\linewidth]{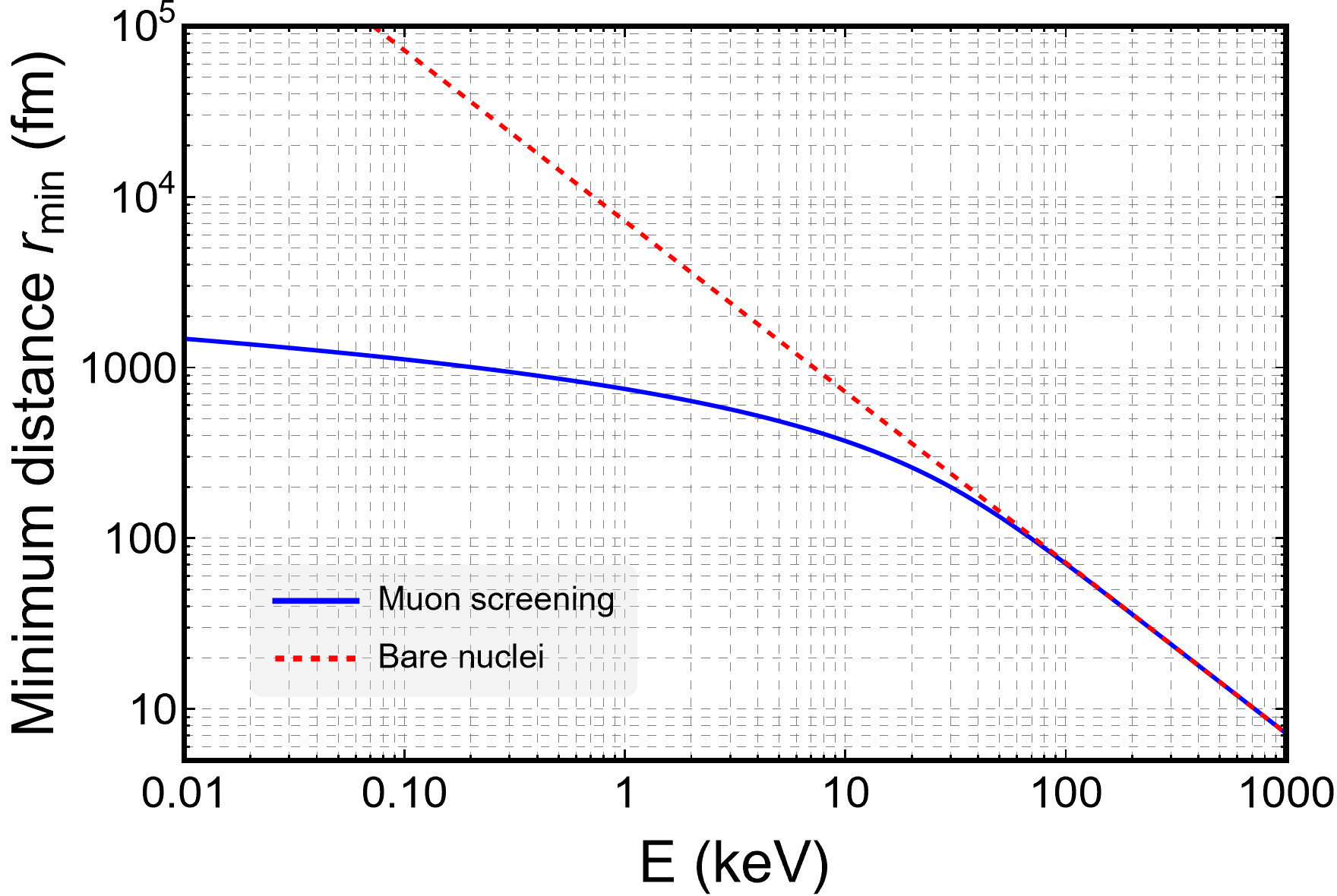}
    \caption{(Color online) Minimum approach distance $r_{\mathrm{min}}$ between p and $^{11}$B as a function of the incident energy $E$. The blue solid curve corresponds to the case with muon screening, while the red dashed curve represents the bare nuclei case.} \label{fig:rvsE}
\end{figure}
By solving $V_{\mathrm{P} \mu-\mathrm{B}}^{\mathrm{eff}}\left(r_{\mathrm{}}\right)=E$, we can calculate the minimum approach distance of $^{11}$B to the proton under muon screening for a given initial energy, as shown in Fig.~\ref{fig:rvsE}; the bare proton case without a muon is also shown as a comparison. It can be seen that in the energy range from 0 to several tens of keV, the inclusion of the muon significantly decreases the minimum distance at the turning point. Beyond about 100 keV, muons almost cannot help anymore, because the incident particles have sufficient energy to enter the region where muon screening is negligible.

\subsection{Penetrability}
\begin{figure}
	\centering
	\includegraphics[width=\linewidth]{muPB_penetrability.png}
    \caption{(Color online) Penetrability $P(E)$ as a function of incident energy $E$. The blue solid curve represents the case with muon screening, while the red dashed curve corresponds to the bare nuclei. Below $E = 33.5$ keV (indicated by the orange dot-dashed line), the blue curve is computed using the WKB approximation. From $E = 33.5$ keV the Airy approximation is employed to smoothly connect to the WKB result. The Airy result intersects with the bare nuclei case at $E = 107.43$ keV (brown dot-dashed line), Above this intersection, the penetrability with muon screening is taken to be identical to that of the bare nuclei, as the muon ceases to enhance penetration.} \label{fig:}
\end{figure}
For energies below 33.5 keV, we use the WKB approximation to calculate the tunneling probability. In this energy region, the inclusion of the muon can enhance the tunneling probability by several orders of magnitude, and the enhancement is more significant at lower incident energies. Starting from 33.5 keV, we use the Airy approximation to connect smoothly and continuously to the previous WKB result. The calculated Airy tunneling probability intersects the tunneling probability for the traditional bare nucleus reaction at an energy of 107.43 keV. We consider that beyond this intersection point, the Airy approximation will also no longer be applicable, because physically the inclusion of the muon should not increase the tunneling probability anymore. Increasing energy only diminishes the enhancement provided by the muon but should not cause the tunneling probability to fall below that of the bare nucleus case. Therefore, beyond this intersection energy, we approximate that the muon no longer enhances the tunneling probability and thus use the traditional exponential form. This aligns with the calculation of the minimum distance: in the low-energy region, the tunneling probability is significantly enhanced by the muon because it greatly reduces the minimum distance between the nuclei. Beyond 100 keV, the muon hardly affects the minimum distance and thus no longer enhances the tunneling probability.

\subsection{Reaction Cross-Section and Reaction Rate}

For the p-$^{11}$B reaction, we adopt a high-precision fitting to the experimental $S(E)$ data:
\begin{equation}
S(E) =
\begin{cases}
S_1(E), & E \leqslant \mathcal{E}_1, \\
S_2(E), & \mathcal{E}_1 < E \leqslant \mathcal{E}_2, \\
S_3(E), & \mathcal{E}_2 < E \leqslant \mathcal{E}_3,
\end{cases}
\end{equation}
where
\begin{equation}
\begin{aligned}
S_1(E) = {} & C_0 + C_1\left(\frac{E}{1\,\mathrm{keV}}\right)
+ C_2\left(\frac{E}{1\,\mathrm{keV}}\right)^2 \\
& + \frac{A_{\mathrm{L}}}{\left(\frac{E - E_{\mathrm{L}}}{1\,\mathrm{keV}}\right)^2
+ \left(\frac{\delta E_{\mathrm{L}}}{1\,\mathrm{keV}}\right)^2},
\end{aligned}
\end{equation}
\begin{equation}
\begin{aligned}
S_2(E) = {} & D_0 + D_1\left(\frac{E - \mathcal{E}_1}{100\,\mathrm{keV}}\right)
+ D_2\left(\frac{E - \mathcal{E}_1}{100\,\mathrm{keV}}\right)^2 \\
& + D_5\left(\frac{E - \mathcal{E}_1}{100\,\mathrm{keV}}\right)^5.
\end{aligned}
\end{equation}
\begin{equation}
S_3(E) = B + \sum_{k=0}^{4}
\frac{A_k}{\left(\frac{E - E_k}{1\,\mathrm{keV}}\right)^2
+ \left(\frac{\delta E_k}{1\,\mathrm{keV}}\right)^2}.
\end{equation}

\begin{table}
\centering
\caption{Fitting parameters for the p-$^{11}$B reaction $S$-factor}
\label{tab:fit_params}

\begin{tabular}{p{0.48\linewidth}p{0.48\linewidth}}
\hline
\\[-21pt] 
\begin{minipage}[t]{\linewidth}
\centering
\vspace{0pt}
\begin{tabular}{lc}
\multicolumn{2}{l}{$E \leq 0.4$ MeV} \\
\midrule
Parameter & Value \\
\midrule
$C_0$ (MeV$\cdot$b) & 197 \\
$C_1$ (MeV$\cdot$b) & 0.240 \\
$C_2$ (MeV$\cdot$b) & $2.31\times10^{-4}$ \\
$A_L$ (MeV$\cdot$b) & $1.82\times10^{4}$ \\
$E_L$ (keV) & 148 \\
$\delta E_L$ (keV) & 2.35 \\
\midrule
\\[2pt]
\cline{1-2} 
\\[-10pt]
\multicolumn{2}{l}{$0.4 < E \leq 0.7$ MeV} \\ 
\midrule
Parameter & Value \\
\midrule
$D_0$ (MeV$\cdot$b) & 330.2 \\
$D_1$ (MeV$\cdot$b) & 102.436 \\
$D_2$ (MeV$\cdot$b) & -58.481 \\
$D_5$ (MeV$\cdot$b) & 0.0933 \\
\end{tabular}
\end{minipage}
&
\begin{minipage}[t]{\linewidth}
\centering
\vspace{0pt}
\begin{tabular}{lc}
\multicolumn{2}{l}{$0.7 < E \leq 10.0$ MeV} \\
\midrule
Parameter & Value \\
\midrule
$B$ (MeV$\cdot$b) & 0.209689 \\
$A_0$ ($\times10^6$ MeV$\cdot$b) & 2.0235 \\
$A_1$ ($\times10^6$ MeV$\cdot$b) & 4.0102 \\
$A_2$ ($\times10^6$ MeV$\cdot$b) & 1.3220 \\
$A_3$ ($\times10^6$ MeV$\cdot$b) & 4.9451 \\
$A_4$ ($\times10^5$ MeV$\cdot$b) & 4.3430 \\
$E_0$ (MeV) & 0.6222 \\
$E_1$ (MeV) & 1.3884 \\
$E_2$ (MeV) & 2.4924 \\
$E_3$ (MeV) & 3.5286 \\
$E_4$ (MeV) & 4.7036 \\
$\delta E_0$ (MeV) & 0.0996 \\
$\delta E_1$ (MeV) & 0.4499 \\
$\delta E_2$ (MeV) & 0.2386 \\
$\delta E_3$ (MeV) & 0.3985 \\
$\delta E_4$ (MeV) & 0.1525 \\
\end{tabular}
\end{minipage}
\\
\\[-9pt] 
\hline
\end{tabular}
\end{table}

Details are explained in Ref.~\cite{wang2026}, and  parameters are listed in Table \ref{tab:fit_params}.

\begin{figure}
	\centering
	\includegraphics[width=\linewidth]{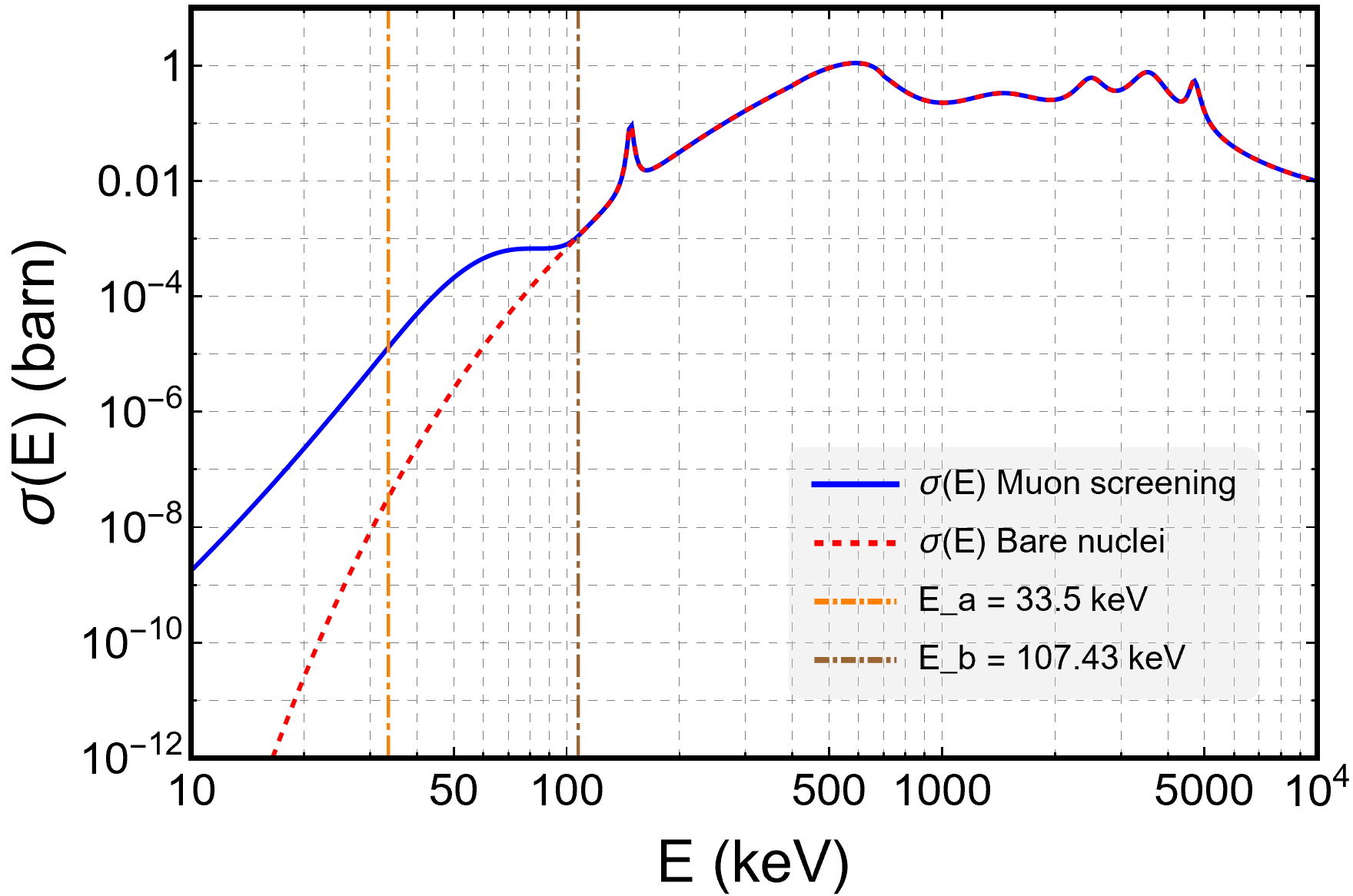}
    \caption{(Color online) Reaction cross section $\sigma(E)$ as a function of incident energy $E$. The blue solid curve represents the cross section with muon screening, while the red dashed curve corresponds to the bare nuclei case. The cross sections are calculated using Eq. (12), with the S-factor taken from Eq. (14).} \label{fig:cs}
\end{figure}
\begin{figure}
	\centering
	\includegraphics[width=\linewidth]{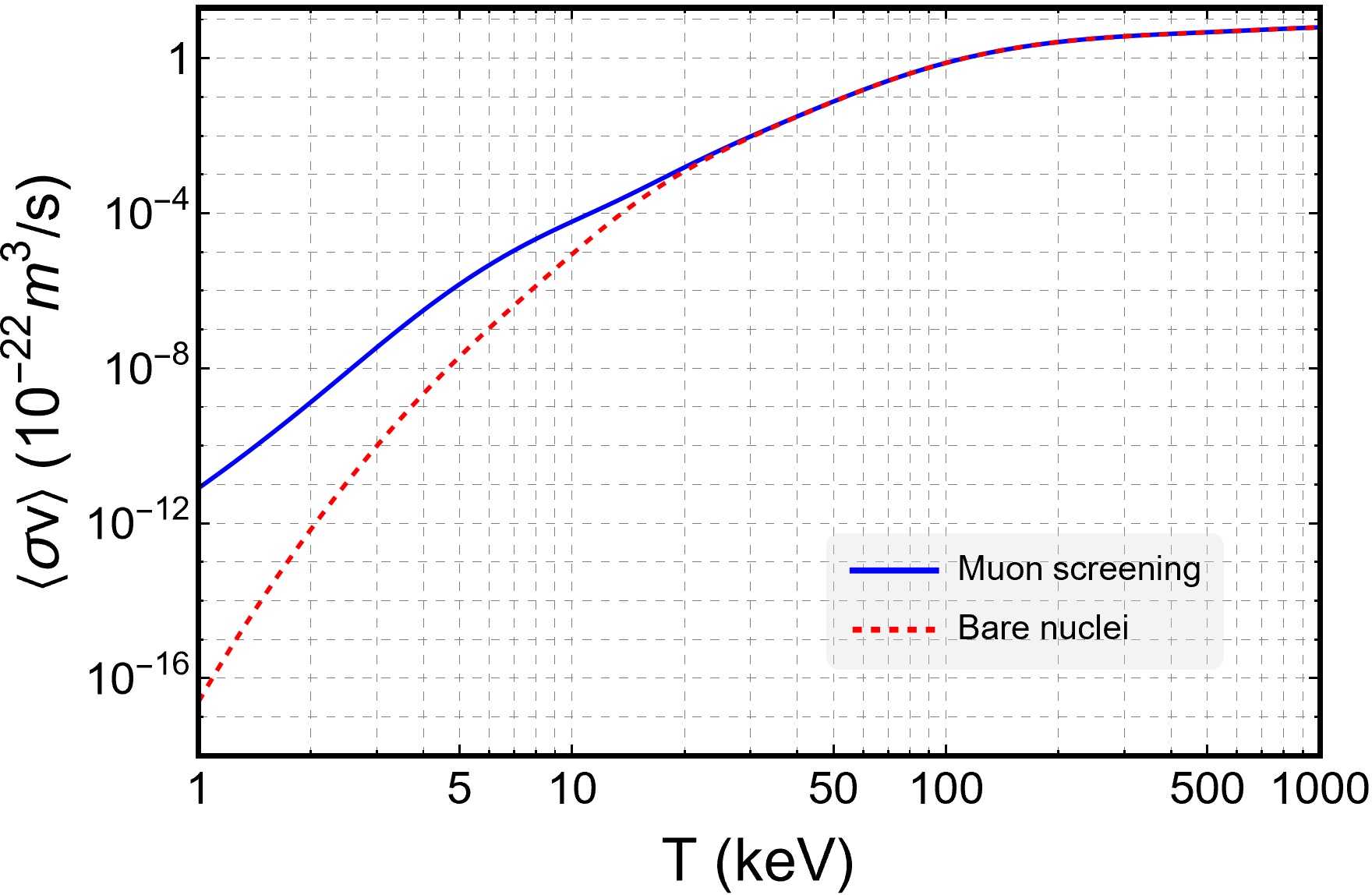}
    \caption{(Color online) Reaction rate $\langle \sigma v \rangle$ as a function of temperature $T$ for the p$\mu$-$^{11}$B system. The blue solid curve represents the rate with muon screening, while the red dashed curve corresponds to the bare nuclei case. The rates are calculated using Eq. (13).} \label{fig:rate}
\end{figure}

In Fig.~\ref{fig:cs}, the blue solid line represents the reaction cross-section with the inclusion of the muon. Similar to the tunneling probability, we can see a significant enhancement in the cross-section before the intersection point at 107.43 keV, but beyond this point, the result remains consistent with the bare nucleus cross-section. The muon does not extend the enhancement of the cross-section to cover the first major resonance peak at 148 keV. And similarly, from the reaction rate results in Fig.~\ref{fig:rate}, it is evident that the muon significantly enhances it before 10 keV, but this enhancement rapidly decays afterward. Beyond 40 keV, the reaction rate fails to show significant enhancement. In magnetic confinement fusion systems for p-$^{11}$B, the energy window of interest for potential net energy gain is approximately between 200-300 keV~\cite{wang2026}.


\section{Conclusion}
\label{conclusion}
Every corner of our lives requires energy. It has been a long-standing dream of humans that one day we can build `artificial suns' on Earth to produce clean and nearly limitless energy. Actually, nuclear fusion energy might be the only primary energy source left in the Universe that we have yet to exploit. 

Scientists have been trying to make fusion reactors since the 1950s, however, so far it's still on the road. Finding breakthroughs requires risk taking, while we believe that revisiting low energy nuclear reactions is a risk worth taking.

Unlike the popular deuterium-tritium (D-T) fusion, proton-boron-11 (p-$^{11}$B) reaction is an attractive aneutronic pathway with many 
advantages, such as abundantly available fuel, primary product ($^4\text{He}$) are charged and can be used for direct energy conversion without the damaging neutron flux, lowers maintenance costs and a much more sustainable end goal, etc.

Current research in $p-^{11}\text{B}$ fusion largely focuses on non-muonic methods, such as laser-driven targets and magnetic confinement, to overcome the high ignition temperature ($>500 \text{ keV}$). In this work, we have investigated a non-thermal route to enhance low-energy p-$^{11}$B fusion by considering a kinetic scenario, in which a muonic hydrogen atom p$\mu$ is first formed and subsequently interacts with a $^{11}$B nucleus. By treating the muon as a static $1s$ screening cloud around the proton, we derived the distance-dependent effective charge and the corresponding modified Coulomb potential experienced by the incident boron nucleus. This muon-induced screening substantially lowers the effective barrier at intermediate separations, thereby increasing the probability of barrier penetration relative to the bare p-$^{11}$B system.

Using the WKB approximation, together with an Airy-function matching procedure in the intermediate regime where the semiclassical description becomes less reliable, we evaluated the penetrability, cross-section, and thermonuclear reactivity of the p$\mu$-$^{11}$B system. The results demonstrate that muon screening can enhance the tunneling probability by several orders of magnitude at sub-100-keV energies, with the strongest effect occurring at the lowest incident energies. Nevertheless, the enhancement rapidly weakens as the incident energy increases, and becomes negligible above the $\sim 100$~keV scale, where the bare Coulomb barrier is already sufficiently reduced by kinetic energy. Consistently, the corresponding enhancement in the reaction cross-section and reactivity is confined to the low-energy regime and does not extend to the dominant resonance region.

These results indicate that the introduction of a muon can substantially reshape the low-energy p-$^{11}$B reaction landscape by dynamically screening the proton charge and reducing the effective Coulomb barrier at intermediate separations. As a consequence, the penetrability, cross-section, and thermonuclear reactivity are all markedly enhanced in the sub-100-keV regime, suggesting that the p$\mu$-$^{11}$B channel may offer a viable route for lowering the reaction threshold and facilitating ignition in aneutronic fusion systems. Although the muon-induced enhancement diminishes as the incident energy increases and becomes negligible in the higher-energy region, the present analysis demonstrates that muon screening can provide a physically meaningful catalytic effect precisely in the low-energy domain where conventional p-$^{11}$B fusion is most strongly suppressed. These findings therefore motivate further exploration of muon-assisted barrier mitigation as a potential ignition mechanism for advanced aneutronic fuels.

\begin{acknowledgments}
This work was supported by Natural Science Foundation of Jiangsu Province (grant no. BK20220122); China Postdoctoral Science Foundation (grant no. 2024M751369); National Natural Science Foundation of China (grant no. 12233002), and Jiangsu Funding Program for Excellent Postdoctoral Talent. Use of the computing resources at \href{www.syli.cloud}{Jiangsu Xilixi Technology} is gratefully acknowledged.
\end{acknowledgments}


\nocite{*}
\bibliographystyle{apsrev4-1}
\bibliography{refs} 
\end{document}